\documentstyle[12pt]{article}

\def\journal{\topmargin .3in	\oddsidemargin .5in
	\headheight 0pt	\headsep 0pt
	\textwidth 5.625in 
	\textheight 8.25in 
	\marginparwidth 1.5in
	\parindent 2em
	\parskip .5ex plus .1ex		\jot = 1.5ex}
\journal

\catcode`\@=11
\def\marginnote#1{}
\newcount\hour
\newcount\minute
\newtoks\amorpm
\hour=\time\divide\hour by60
\minute=\time{\multiply\hour by60 \global\advance\minute by-\hour}
\edef\standardtime{{\ifnum\hour<12 \global\amorpm={am}%
	\else\global\amorpm={pm}\advance\hour by-12 \fi
	\ifnum\hour=0 \hour=12 \fi
	\number\hour:\ifnum\minute<10 0\fi\number\minute\the\amorpm}}
\edef\militarytime{\number\hour:\ifnum\minute<10 0\fi\number\minute}
\def\draftlabel#1{{\@bsphack\if@filesw {\let\thepage\relax
   \xdef\@gtempa{\write\@auxout{\string
      \newlabel{#1}{{\@currentlabel}{\thepage}}}}}\@gtempa
   \if@nobreak \ifvmode\nobreak\fi\fi\fi\@esphack}
	\gdef\@eqnlabel{#1}}
\def\@eqnlabel{}
\def\@vacuum{}
\def\draftmarginnote#1{\marginpar{\raggedright\scriptsize\tt#1}}
\def\draft{\oddsidemargin -.5truein
	\def\@oddfoot{\sl preliminary draft \hfil
	\rm\thepage\hfil\sl\today\quad\militarytime}
	\let\@evenfoot\@oddfoot	\overfullrule 3pt
	\let\label=\draftlabel
	\let\marginnote=\draftmarginnote
   \def\@eqnnum{(\theequation)\rlap{\kern\marginparsep\tt\@eqnlabel}%
\global\let\@eqnlabel\@vacuum}  }
\def\preprint{\twocolumn\sloppy\flushbottom\parindent 2em
	\leftmargini 2em\leftmarginv .5em\leftmarginvi .5em
	\oddsidemargin -.5in	\evensidemargin -.5in
	\columnsep .4in	\footheight 0pt
	\textwidth 10in	\topmargin  -.4in
	\headheight 12pt \topskip .4in
	\textheight 7.1in \footskip 0pt
	\def\@oddhead{\thepage\hfil\addtocounter{page}{1}\thepage}
	\let\@evenhead\@oddhead	\def\@oddfoot{}	\def\@evenfoot{} }
\def\numberbysection{\@addtoreset{equation}{section}
	\def\theequation{\thesection.\arabic{equation}}}
\def\underline#1{\relax\ifmmode\@@underline#1\else
	$\@@underline{\hbox{#1}}$\relax\fi}
\def\titlepage{\@restonecolfalse\if@twocolumn\@restonecoltrue\onecolumn
     \else \newpage \fi \thispagestyle{empty}\c@page\z@
	\def\thefootnote{\fnsymbol{footnote}} }
\def\endtitlepage{\if@restonecol\twocolumn \else \newpage \fi
	\def\thefootnote{\arabic{footnote}}
	\setcounter{footnote}{0}}  
\catcode`@=12
\relax
\def\figcap{\section*{Figure Captions\markboth
	{FIGURECAPTIONS}{FIGURECAPTIONS}}\list
	{Figure \arabic{enumi}:\hfill}{\settowidth\labelwidth{Figure 999:}
	\leftmargin\labelwidth
	\advance\leftmargin\labelsep\usecounter{enumi}}}
 \relax
\def\tablecap{\section*{Table Captions\markboth
	{TABLECAPTIONS}{TABLECAPTIONS}}\list
	{Table \arabic{enumi}:\hfill}{\settowidth\labelwidth{Table 999:}
	\leftmargin\labelwidth
	\advance\leftmargin\labelsep\usecounter{enumi}}}
 \relax
\def\reflist{\section*{References\markboth
	{REFLIST}{REFLIST}}\list
	{[\arabic{enumi}]\hfill}{\settowidth\labelwidth{[999]}
	\leftmargin\labelwidth
	\advance\leftmargin\labelsep\usecounter{enumi}}}
 \relax
\makeatletter
\newcounter{pubctr}
\def\publist{\@ifnextchar[{\@publist}{\@@publist}}
\def\@publist[#1]{\list
	{[\arabic{pubctr}]\hfill}{\settowidth\labelwidth{[999]}
	\leftmargin\labelwidth
	\advance\leftmargin\labelsep
	\@nmbrlisttrue\def\@listctr{pubctr}
	\setcounter{pubctr}{#1}\addtocounter{pubctr}{-1}}}
\def\@@publist{\list
	{[\arabic{pubctr}]\hfill}{\settowidth\labelwidth{[999]}
	\leftmargin\labelwidth
	\advance\leftmargin\labelsep
	\@nmbrlisttrue\def\@listctr{pubctr}}}
 \relax
\makeatother
\catcode`\@=11
\def\section{\@startsection {section}{1}{0pt}{-3.5ex plus -1ex minus
 -.2ex}{2.3ex plus .2ex}{\raggedright\large\bf}}
\catcode`\@=12
\newskip\humongous \humongous=0pt plus 1000pt minus 1000pt
\def\caja{\mathsurround=0pt}

\newif\ifdtup
\def\panorama{\global\dtuptrue \openup1\jot \caja
	\everycr{\noalign{\ifdtup \global\dtupfalse
	\vskip-\lineskiplimit \vskip\normallineskiplimit
	\else \penalty\interdisplaylinepenalty \fi}}}
\def\eqalignno#1{\panorama \tabskip=\humongous
	\halign to\displaywidth{\hfil$\displaystyle{##}$
	\tabskip=0pt&$\displaystyle{{}##}$\hfil
	\tabskip=\humongous&\llap{$##$}\tabskip=0pt
	\crcr#1\crcr}}

\def\oldreffmt#1{\rlap{[#1]} \hbox to 2\parindent{}}

\def\figfmt#1{\rlap{Figure {#1}} \hbox to 1in{}}

\let\vev\VEV

\def\abs#1{\left| #1\right|}

\def\beq{\begin{equation}}
\def\eeq{\end{equation}}

\def\bea{\begin{eqnarray}}

\def\eea{\end{eqnarray}}
\catcode`\@=11
\def\eqnarray{\stepcounter{equation}\let\@currentlabel=\theequation
\global\@eqnswtrue
\global\@eqcnt\z@\tabskip\@centering\let\\=\@eqncr
\gdef\@@fix{}\def\eqno##1{\gdef\@@fix{##1}}%
$$\halign to \displaywidth\bgroup\@eqnsel\hskip\@centering
  $\displaystyle\tabskip\z@{##}$&\global\@eqcnt\@ne
  \hskip 2\arraycolsep \hfil${##}$\hfil
  &\global\@eqcnt\tw@ \hskip 2\arraycolsep $\displaystyle\tabskip\z@{##}$\hfil
   \tabskip\@centering&\llap{##}\tabskip\z@\cr}

\def\@@eqncr{\let\@tempa\relax
    \ifcase\@eqcnt \def\@tempa{& & &}\or \def\@tempa{& &}
      \else \def\@tempa{&}\fi
     \@tempa \if@eqnsw\@eqnnum\stepcounter{equation}\else\@@fix\gdef\@@fix{}\fi
     \global\@eqnswtrue\global\@eqcnt\z@\cr}

\catcode`\@=12
\font\tenbifull=cmmib10 
\font\tenbimed=cmmib10 scaled 800
\font\tenbismall=cmmib10 scaled 666
\relax

\def\thefootnote{\fnsymbol{footnote}}
\begin{document}
\begin{titlepage}
\begin{center}
\today          \hfill     LBL-38065 \\
          \hfill    UCB-PTH-95/44 \\

\vskip .5in

{\large \bf Predictions From A U(2) Flavour Symmetry
\\ In Supersymmetric Theories.}
\footnote{This work was supported in part by the Director, Office of
Energy Research, Office of High Energy and Nuclear Physics, Division of
High Energy Physics of the U.S. Department of Energy under Contract
DE-AC03-76SF00098 and in part by the National Science Foundation under
grant PHY-90-21139.}

\vskip .3in
{\bf Riccardo Barbieri}

{\em Physics Department\\
University of Pisa and INFN Sez. di Pisa\\
I-56126 Pisa, Italy}

\medskip

{\bf George Dvali}

{\em CERN\\
    Geneva\\
         Switzerland}

\medskip

and\\

\medskip

{\bf Lawrence J. Hall}

{\em Theoretical Physics Group, LBNL, and\\
Physics Department, University of California\\
    Berkeley, California 94720}
\end{center}

\vskip .3in

\begin{abstract}
In a generic supersymmetric extension of the Standard Model,
whether unified or not, a simple and well motivated U(2) symmetry,
acting on the lightest two generations, completely solves the
flavour changing problem and necessarily leads to a predictive texture
for the Yukawa couplings.

\end{abstract}
\end{titlepage}
\renewcommand{\thepage}{\roman{page}}
\setcounter{page}{2}
\mbox{ }

\vskip 1in

\begin{center}
{\bf Disclaimer}
\end{center}

\vskip .2in

\begin{scriptsize}
\begin{quotation}
This document was prepared as an account of work sponsored by the United
States Government. While this document is believed to contain correct
 information, neither the United States Government nor any agency
thereof, nor The Regents of the University of California, nor any of their
employees, makes any warranty, express or implied, or assumes any legal
liability or responsibility for the accuracy, completeness, or usefulness
of any information, apparatus, product, or process disclosed, or represents
that its use would not infringe privately owned rights.  Reference herein
to any specific commercial products process, or service by its trade name,
trademark, manufacturer, or otherwise, does not necessarily constitute or
imply its endorsement, recommendation, or favoring by the United States
Government or any agency thereof, or The Regents of the University of
California.  The views and opinions of authors expressed herein do not
necessarily state or reflect those of the United States Government or any
agency thereof, or The Regents of the University of California.
\end{quotation}
\end{scriptsize}

\vskip 2in

\begin{center}
\begin{small}
{\it Lawrence Berkeley Laboratory is an equal opportunity employer.}
\end{small}
\end{center}

\newpage
\renewcommand{\thepage}{\arabic{page}}
\setcounter{page}{1}
\noindent{\bf 1. Introduction and Motivation}

\medskip

        The accomplishment of the electroweak precision tests in the first
phase of LEP \cite{Renton},
showing a remarkable agreement between the experimental
results and the expectation of the Standard Model, has provided
indirect evidence for the Higgs picture of the electroweak symmetry breaking.
This brings the focus, more than ever, to the ``fermion mass problem" of the
SM: the inelegant description of fermion masses and mixings in terms of a
number of arbitrary dimensionless Yukawa couplings of the fermions to the
Higgs boson. These couplings show a strong hierarchical pattern, with only one
coupling of order unity, largely dominating over the others, for which
the SM provides no understanding.

        Along an independent line of consideration,  any evidence for the
existence of the Higgs boson, as the one provided by LEP, strengthens the
view that supersymmetry may be a relevant symmetry of nature, as realized,
for example, in the Minimal Supersymmetric Standard Model. It is well
known, on the other hand, that the ``fermion mass problem", as defined above,
appears in the MSSM precisely in the same way as in the SM itself. At the
same time, however, it is also well known that the description of flavour
in the MSSM has a special feature that distinguishes it from the SM
\cite{DG,EN}. In the
SM, the description of flavour in terms of Yukawa couplings provides a neat
solution of the ``flavour-changing" problem: the flavour changing neutral
current processes are automatically suppressed to the required
phenomenological level via the GIM mechanism \cite{GIM}.
On the contrary, for this to happen
in the MSSM, appropriate universality assumptions for the soft
supersymmetry breaking terms are also required as an independent input.

In supersymmetric theories the fermion mass and flavour-changing problems are
different aspects of a single ``flavour" problem. An interesting approach is to
study flavour symmetries which simultaneously address both aspects
\cite{DLK}.
The purpose of this paper is to show that a simple,
well-motivated flavour symmetry,
acting on the lightest two generations, completely solves the flavour-changing
problem, and necessarily leads to a predictive texture for the Yukawa
matrices.

There are many candidate flavour-symmetry groups $G_f$, each having several
distinct symmetry breaking patterns \cite{FlSym}.
In general, $G_f$ must be contained in the full
global symmetry group of the SM in the limit of vanishing Yukawa couplings,
$U(3)^5$, with each $U(3)$ acting in the 3-dimensional generation space and,
independently, on one of the five irreducible representations under the
``vertical" gauge group, $(Q, u^c, d^c, L, e^c)$,
which compose the usual 15-plet
of matter fields per generation. Although our considerations, unless
otherwise stated, do not depend on assuming a unified symmetry in the
vertical direction, we nevertheless choose to restrict our attention to
schemes which might be applicable in such a case too. In particular, if we
consider the case of full unification of
$$
\psi = (Q, u^c, d^c L, e^c)\eqno(1)
$$
into a single representation of the gauge group, as, e.g., in the case of
$SO(10)$, we are lead to consider $U(3)$ as the maximal possible
$G_f$. On the other
hand, the large Yukawa coupling of the top quark, $\lambda_t$,
represents a violent
breaking of this family symmetry, which will reflect itself also in the
sfermion spectrum, in fact both in the squarks and in the sleptons in the
case of a unified theory \cite{BH}. For this reason, although the large
$\lambda_t$  might also
result from the spontaneous breaking of the full $U(3)$ symmetry, we will
mostly consider, in the following, a $U(2)$ family symmetry, under which the
vertical multiplets $\psi_i$, $i=1,2,3$ for the 3 families, transform as a
$\underline{2} + \underline{1}$ representation \cite{PT}.

        An independent argument for considering only a $U(2)$ rather than the
larger $U(3)$ symmetry is the following. Suppose, for definiteness, in the
physical basis both for fermions and sfermions, that the mixing matrices in
the gaugino-matter interactions are close to the standard CKM matrix. In
such a case, a splitting between the masses of the first two generations of
sfermions, of given charge, comparable to their mean mass leads to a serious
flavour-changing problem, e.g., in $K^0 - \bar{K}^0$
mixing or in the rate for $\mu \rightarrow e \gamma$ or, if
physical phases are present, in the CP violating $\epsilon$
parameter in $K$ physics.
By a related phenomenon, a large electric dipole moment for the electron
and/or the neutron can also be generated. We will show how such problems
can be taken care of by an appropriately broken $U(2)$ symmetry. On the
contrary, a splitting between the third and the first two generations of
sfermions is not necessarily a problem. It has actually been shown in
minimal unified theories that the splitting produced in the sfermion masses
as a consequence of the large top Yukawa coupling gives rise to very
interesting signatures in lepton flavour violating processes
\cite{BH} and in EDMs of the
electron and the neutron \cite {DH,BH}.

Hence we are led to consider $G_f = U(2)$, realizing that it is likely that
this is a remnant of a strongly broken $U(3)$ flavour symmetry, the maximal
flavour group for a unified theory.

\medskip

\noindent{\bf 2. The model}

\medskip

        We first consider a set of general assumptions, which are:

\begin{enumerate}
        \item[i)] The flavour group is $G_f = U(2)$,
under which the three generations $\psi_i = \psi_a, \psi$, with $a=1,2$,
transform as $\underline{2} + \underline{1}$.

        \item[ii)] The Higgs field(s) $H$ transform as some representation,
reducible or irreducible, of the vertical gauge group $G$, but are pure $G_f$
singlets.

       \item[ iii)]  The flavour group $G_f$, which has rank 2,
is broken by two vacuum expectation values: of a doublet $\phi^a$
and of a singlet, or, more
precisely, a 2-index antisymmetric tensor $\phi^{ab}$, such that, without loss
of generality
\end{enumerate}
$$
\vev{\phi^a} = \pmatrix{0 \cr V}, \;  \; \vev{\phi^{ab}} =
v\epsilon^{ab}\eqno(2a)
$$
For these vevs we assume the hierarchy $V \gg v$, so that
$$
U(2)\; {\stackrel{V}{\longrightarrow}}\; U(1)\; {\stackrel{v}{\longrightarrow}}
\; \mbox{nothing}.\eqno(2b)
$$

The most general $G_f$ invariant superpotential relevant for generating fermion
masses, $W_Y$, linear in $H$ and bilinear in the matter fields, with
non-renormalizable terms weighted by inverse powers of a mass scale
$M \gg V \gg v$, is
$$
W_Y=\psi \lambda_1 H \psi + {\phi^a\over M} \psi \lambda_2 H \psi_a +
{\phi^{ab}\over M} \psi_a\lambda_3H\psi_b +{\phi^a\phi^b\over M^2}
\psi_a\lambda_4 H\psi_b\eqno(3)
$$
At scale $M$ we assume that the vertical gauge symmetry is reduced to the
SM gauge group, so that, in general
$$
\eqalignno{
\psi_i \lambda H\psi_j &= \lambda_U Q_i u^c_j h_2 + \lambda'_U u^c_i Q_j h_2
                        + \lambda_D Q_i d^c_j h_1\cr
                       &+ \lambda'_D d^c_i Q_j h_1 + \lambda_L L_i e^c_j h_1
                        + \lambda'_L e^c_i L_j h_1&(4)\cr}
$$

For reasons that will become clear in the next Section, rather
than considering the most general $W_Y$, we require that the non-renormalizable
interactions in the superpotential (3) be generated from a
renormalizable superpotential by integrating out a heavy family,
$\chi^a + \bar{\chi}_a$,
vector-like under the vertical gauge group, transforming as a doublet
under the flavour group. In full generality, such a superpotential is
$$
W=\psi\lambda H\psi + \psi_a \lambda' H \chi^a + \phi^{ab} \psi_a \sigma
\bar{\chi}_b + \chi^a M\bar{\chi}_a + \phi^a \psi \tau \bar{\chi}_a\eqno(5)
$$
Here, as in eq. 3, there is an implicit vertical structure for every
term, which is left understood.

By integrating out the heavy fields $\chi^a + \bar{\chi}_a$,
all terms of the superpotential (3) are reproduced, except the last one.
In turn, this superpotential, after insertion of the vevs (2) , leads to
the following texture of the Yukawa couplings for the Up quarks, the Down
quarks and the charged Leptons
$$
\lambda^{U,D,L} = \pmatrix{ O&d&O\cr
                            -d&O&b\cr
                            O&c&a}^{U,D,L}\eqno(6)
$$
where, setting $V/M = \epsilon$ and $v/M = \epsilon'$,
$$
a=O(1) \; \;  b,c = O(\epsilon) \; \; d=O(\epsilon ').\eqno(7)
$$
By an approximate diagonalization of these Yukawa couplings, taking into
account the hierarchy in the mass eigenvalues, it is a simple matter to
show that the CKM matrix takes the following form \cite{DHR}
$$
V_{CKM} = \pmatrix{ 1& s^D_{12}+ s^U_{12} e^{-i\phi} & s^U_{12} s_{23}\cr
                    -s^U_{12}-s^D_{12} e^{-i\phi} & e^{-i\phi} & s_{23}\cr
                    s^D_{12} s_{23}& -s_{23} & e^{i\phi}}\eqno(8)
$$
where
$$
s^U_{12} = \sqrt{ {m_u\over m_c}}, \ \ s^D_{12} = \sqrt{ {m_d\over m_s}}.
\eqno(9)
$$
As a consequence, if we stick to the usual current algebra determination of
the light quark masses\cite{GL}, barring in particular $m_u = 0$ \cite{L},
this gives the relation
$$
\abs{ {V_{ub}\over V_{cb}} }= \sqrt{ {m_u\over m_c} } = 0.061 \pm
0.009,\eqno(10a)
$$
to be compared with the current world average \cite{RPP}
$$
\abs{ {V_{ub}\over V_{cb}} }_{exp} = 0.08 \pm 0.02. \eqno(10b)
$$
Furthermore, it predicts
$$
\abs{ {V_{td}\over V_{ts}}} = \sqrt{ {m_d\over m_s}} = 0.226 \pm
0.009,\eqno(11)
$$
against a current range of 0.1 to 0.3, and, to account for the observed
value of $|V_{us}| = 0.221 \pm 0.002$,
it also predicts a large CP violating CKM phase, $\sin \phi \ge 0.9$.

How general is the form (8,9) that we have obtained
for the CKM matrix? It is easy to see that the same form would have been
obtained from the most general $W_Y$, since, in this case, the texture of the
fermion mass matrices acquires also a non vanishing 22 entry, which does
not affect eq. (8,9) \cite{DHR}.
Hence, a CKM matrix of the form (8,9), leading to
predictions (10a, 11), results from an arbitrary theory at scale $M$, provided
only that it possesses a $U(2)$ flavour symmetry broken below $M$ by the two
vevs
of (2a). While this is a remarkably general result, the ``renormalizable" model
is of particular interest since it is the simplest viable U(2) invariant theory
at scale $M$, and it possesses a
very distinctive flavour and CP violating behaviour, induced by
the gaugino-matter interactions.

The exchange of one heavy family,
$\chi + \bar{\chi}$, singlet instead of doublet under $G_f$,
would not have given rise to a
realistic texture. On the contrary, as we are also going to show, a set of
heavy matter fields transforming under $G_f$ as $\underline{2} +
\underline{1}$, $\chi + \bar{\chi} + \chi^a + \bar{\chi}_a$, maintains the same
properties of the pure heavy doublet model provided
the coupling $\psi_a \phi^a \bar{\chi}$
in the superpotential is forbidden by an appropriate extension of the
flavour symmetry.

\medskip

\noindent{\bf 3. Sfermion masses}

\medskip

As anticipated, the $U(2)$ flavour symmetry not only constrains the
form of the Yukawa couplings, but also the amount of non-degeneracy between
the sfermions of the first and second generation. In turn, this is the key
quantity which affects the ``flavour changing problem" in a supersymmetric
theory.

Supersymmetry breaking is assumed to occur as in supergravity \cite{BFS}, with
no universality-type constraint on the scalar masses or on the analytic
terms, both characterized in the usual way by a scale $m \ll M$ .
In the $U(2)$ invariant limit, the scalar mass squared matrices for each charge
are diagonal with the first two entries degenerate. Consider the general $U(2)$
invariant theory, based on the assumptions i) - iii), with $F$ terms given by
(3). Since $\phi$ appears only in the non-renormalizable terms, as $M
\rightarrow \infty$ the effects of $U(2)$ breaking decouple. The same holds
true
for the $D$ terms where all $U(2)$ breaking is again suppressed by powers of
$M$, for example, as in the soft operator: $m^2 \psi^* \psi_a \phi^a /M$.
Hence the deviation of the scalar mass matrix from the $U(2)$ invariant form
is described by entries involving powers of $\epsilon$
and $\epsilon'$, the same small parameters that generated the hierarchies in
the fermion mass matrices. In particular, because $U(2)$ is kept as a good
symmetry beneath the scale $M$ of any new heavy particles having renormalizable
interactions with the light matter, the potentially dangerous radiative
corrections to universality \cite{HKR} are suppressed by powers of $\epsilon$
and $\epsilon'$. From a general operator analysis one finds that
the sfermions of the first and second generations are split by a
relative amount of order $\epsilon^2$,
so that the typical supersymmetry breaking scalar masses of the
i-th generation, $m_{Si}$, are related to the i-th generation
fermion masses, $m_{Fi}$, by
$$
{m^2_{S1} - m^2_{S2}\over m^2_{S1} + m^2_{S2}} = O
\left( {m_{F2}\over m_{F3}}\right)\eqno(12)
$$
leading to a problematic contribution to the $\epsilon$ parameter of kaon
physics \cite{PT}.

In the ``renormalizable" model of (5), the $U(2)$ flavour symmetry is broken by
interactions of $\phi^a$ and $\phi^{ab}$ coupling light and heavy generations.
Any such mass mixing effect generates at tree level both fermion mass
hierarchies \cite{FN} and deviations from the flavour symmetric form of the
scalar mass matrices \cite{DP}. The latter effect is in general a powerful
constraint on supersymmetric theories of fermion masses which use the
Froggatt-Nielsen mechanism, and is more dangerous than the radiative effects
of \cite{HKR}. However, in the present case, with the exchange of only a heavy
$U(2)$ doublet generation, a non-degeneracy between the scalar masses of the
first two generations is induced only at order $\epsilon^2 \epsilon'^2$ , or
$$
{m^2_{S1}-m^2_{S2}\over m^2_{S1} + m^2_{S2} } =
 O\left( {m_{F1}m_{F2}^2 \over m^3_{F3}}\right)\eqno(13)
$$
making any effect in flavour and/or CP violations due to the splitting
between $m_{S1}$ and  $m_{S2}$ completely negligible. This arises because
$\epsilon$ is the only
SU(2) breaking parameter and because, in the limit of vanishing $\epsilon'$,
corresponding to massless fermions of the first  generation, the exact
composition of the heavy states, as determined by the superpotential of eq.
(5) does not contain the light doublets $\psi_a$ at all.  This means that, as
$\epsilon' \rightarrow 0$,
in the most general supersymmetry breaking potential (apart from terms
linear in $H$)
$$
\eqalignno{
V&= m^2_1 \abs{\psi_a}^2 + m^2_2 \abs{ \chi^a}^2 + m^2_3 \abs{\psi}^2 + m^2_4
\abs{\bar{\chi}_a}^2\cr
&+ \phi^{ab} \psi_aA_1\lambda_1\bar{\chi}_b + \chi^bA_2 M\bar{\chi}_b + \phi^b
\psi A_3 \sigma \bar{\chi}_b&(14)\cr}
$$
the scalar components of the light first two generations do not feel at all
the breaking of the $U(2)$ symmetry and therefore remain exactly degenerate.
An explicit calculation, including the $\epsilon'$-terms,
shows that the corrections to exact degeneracy are of
order $\epsilon^2 \epsilon'^2$.  Precisely this same argument
can be repeated in the case of heavy matter states consisting of a
$\underline{2} + \underline{1}$
representation under $G_f$. In this case, the decoupling of the light doublets
as $\epsilon'$ goes to zero requires that the coupling $\psi_a \phi^a
\bar{\chi}$  be forbidden by an appropriate extension of the flavour symmetry.

\medskip

\noindent{\bf 4. Flavour and CP violations}

\medskip

        As particularly important characteristic observables, we consider
the rate for the decay $\mu \rightarrow e \gamma$,
the CP violating parameter $\epsilon$ in $K$ physics and the
electric dipole moments for the electron and/or the neutron. For given
values of the mixing angles and
particle masses, the various contributions to these observables from one
loop supersymmetric particle exchanges have been computed, e.g., in reference
\cite{BH}.

As mentioned, in the ``renormalizable" model, no sizeable
contribution to these observables is expected from the exchanges of the
highly degenerate sfermions of the first and second generation. On the
other hand, a calculation of the contribution from the exchange of the
third generation sfermions would require knowing the splitting of their
mass with respect to that of the first two generations, which
is not determined by pure symmetry arguments. Still the pattern of masses
characteristic of this model allows several precise considerations to be
made.

        Let us work in the superfield basis where the sfermion squared mass
matrices are diagonal. In this basis, the fermion masses are diagonalized
as usual by
$$
M^{U,D,L} = V^{U,D,L}_\ell M^{U,D,L}_{diag} V^{\dagger U,D,L}_r\eqno(15)
$$
where the matrices $V_\ell$ and $V_r$ define
the mixing matrices in the gaugino-matter
interaction vertices. Taking into account the texture of (6), it is
immediate to see that $V_\ell$ and $V_r$ have the following approximate form
$$
V^{U,D,L}_{\ell,r} = \pmatrix{ 1&s^{\ell,r}_{12} & O\cr
                              -s^{\ell,r}_{12} & 1 & s^{\ell,r}_{23}\cr
                              s^{\ell,r}_{12}s^{\ell,r}_{23}&-s^{\ell,r}_{23}
& 1}^{U,D,L}
\eqno(16)
$$
where
$$
(s^\ell_{23}s^r_{23})^{U,D,L} = \left({m_2\over m_3}\right)^{U,D,L} \; \; \;
-s^{\ell^{U,D,L}}_{12} = s^{r^{U,D,L}}_{12} = \sqrt{ {m_1\over m_2}}^{U,D,L}
\eqno(17)
$$
and all phases have been eliminated by phase redefinitions of the superfields
$u, u^c, d, d^c, e, e^c$.
That this is possible at all  is a non trivial property of the texture (6).
The implication of this for CP violation is discussed below.

        In general, the theory contains two sources of flavour violations:
the mixing matrices $V_\ell$ and $V_r$ and the $A$-terms linear in $H$,
whose general form, before integrating out the heavy fields, $\chi^a +
\bar{\chi}_a$ is
$$
\psi A \lambda H \psi + \psi_a A'\lambda' H \chi^a \eqno(18)
$$
In turn, the $\mu \rightarrow e \gamma$
decay amplitude receives a contribution from the mixing
matrices in the gaugino-higgsino interactions and another from the $A$-terms.
The first one is proportional to
$V_{\ell_{31}}^L m_\tau V_{r_{32}}^L$, or
$V_{r_{31}}^L m_\tau V_{\ell_{31}}^L$, both determined, from eq.(17), to
be equal to $\sqrt{m_e m_\mu}$, which is a factor of ten bigger
than $V_{td}^{CKM} m_\tau V_{ts}^{CKM}$, as arises in minimal $SO(10)$
\cite{BH}. Leaving aside the $A$-term contributions, which
contain one unknown parameter for any independent Yukawa coupling,
the $\mu \rightarrow e \gamma$
decay rate, for $m_{gaugino} = m_{sleptons} = m$, is estimated as
$$
BR(\mu\to e \gamma ) = O(10^{-10}) \left({300 GeV\over m}\right)^4\left(
{\Delta_\ell\over m^2}\right)^4\eqno(19)
$$
where $\Delta_\ell = m^2_{\tilde{\tau}} -  m^2_{\tilde{\mu}}$
is the splitting between the stau and smuon masses.

        Let us turn now to CP violation and consider first the leptons. As
aforementioned, all phases can be eliminated from the mass matrix by
redefining the fields $L$ and $e^c$. We assume that the only source of CP
violation is in the Yukawa couplings, so that the A parameters, the
gaugino masses and the $\mu$-term are all real. This means that the same
redefinition of the $L$ and $e^c$ superfields which makes the mass matrix real,
also makes real the $A$-terms arising from eq. (18). Under the stated
assumptions, there is no CP violation in the lepton sector: in particular
there is no one loop contribution from gaugino-slepton exchanges to the
electric dipole moment of the electron.

        At variance with the lepton case, the phases of the quark mass
matrices can only be eliminated by independent redefinitions of the $u$ and
$d$ fields, which of course explains why CP is violated, as shown by the
presence of the physical phase in the CKM matrix, eq. (8). Nevertheless CP
violation is more ``screened" than in the generic case.

        Let us consider first the dipole moments, electric or
chromoelectric, (EDM), of the quarks. In all the one loop diagrams with
internal squarks possibly contributing to the EDM of the up or the down
quarks, only the mass terms or the $A$-terms involving the $u^c$ or the $d^c$
are respectively relevant. It is therefore possible, by redefining the
superfields $Q$ and $u^c$, or $Q$ and $d^c$, to rotate away all phases both
from the
mass terms and from the supersymmetric one loop dipole moments for the up
and the down quarks. As in the lepton case, there is no one loop EDM for
the quarks too.

        By working in the basis where the d-quark mass has been made real,
it is also clear that the supersymmetric box-diagram contribution to the
$\Delta S = 2$ left-right effective Hamiltonian operator
$(\bar{s}_L \gamma_\mu d_L) (\bar{s}_R \gamma_\mu d_R)$
has a real coefficient. In this
same basis, however, the conventional $\Delta S = 1$ operator
$(\bar{s}_L \gamma_\mu d_L) (\bar{u}_L \gamma_\mu u_L)$ has a complex
coefficient from the CKM matrix element $V_{us}$ of eq. (8), which induces an
$\epsilon$
parameter. Taking into account that the gluino exchange contribution to the
$\Delta S = 2$ operator is proportional to
$$
 \left(V^D_{\ell_{31}}V^D_{r_{32}}\right)^2 = {m_dm_s\over m_b^2},\eqno(20)
$$
a factor of about one hundred times bigger than
$(V_{td}^{CKM} V_{ts}^{CKM})^2$, which would arise in minimal $SO(10)$
\cite{BH}, this leads to
$$\epsilon = O(10^{-2}) \left({1 TeV\over m}\right)^2\left({\Delta_q\over
m^2}\right)^2
\sin\phi\eqno(21)
$$
for $m_{gluino} = m_{squark} = m$,
where $\Delta_q = m^2_{\tilde{b}} - m^2_{\tilde{s}}$
is the splitting between the relevant squark masses. Recall
that the observed value of $V_{us}$ suggests a large value for the CP violating
phase, or $\abs{\sin \phi} \approx 1$.

\medskip

\noindent{\bf 5. Conclusions}

\medskip

In this paper we have described a supersymmetric model which solves
the ``flavour-changing problem" by virtue of a spontaneously broken,
non-Abelian
flavour symmetry and, at the same time, forces an interesting
texture of fermion masses which leads to some predictions for the CKM
parameters. The masses of the sfermions of the first two generations are
highly degenerate, resulting in a strong suppression of the related
contributions to the flavour and/or CP violating observables.

        A detailed prediction of all such observables would require knowing
the splitting between the third and the first two generations of squarks
and sleptons, which is not fixed by pure symmetry considerations. In any
case, under the stated assumptions, we can say that we do not expect
sizeable EDMs for the quarks or the electron. On the contrary, both the
$\mu \rightarrow e \gamma$ decay rate and the
$\epsilon$ parameter in $K$ physics receive significant
contributions from supersymmetric one loop diagrams if the splitting
between the third and the first two generations of squarks and sleptons,
$\Delta_{q,l}$, is indeed sizeable.
In a unified theory, the maximal flavour symmetry, $U(3)$, is necessarily
strongly broken to $U(2)$ by the large top quark Yukawa coupling, and we know
of no mechanism which is able to protect a small scalar mass splitting
$\Delta_{q,l}$.
In establishing the
relative importance of $\mu \rightarrow e \gamma$ and $\epsilon$,
the family independent effect of the gluino on the squark masses may
play a role \cite{BH}.

In previous papers, two of us have shown that \cite{BH}, in a unified theory,
the large top Yukawa
coupling is indeed a source  of significant splitting
between the masses of the third generation sleptons and squarks with
respect to the first and second generation sfermions of the same charge. In
computing the induced flavour and CP violations, the effect of a possible
splitting between the first and second generation sfermions was ignored
there. As pointed out in the introduction, such an effect, if indeed
existing, would have been much bigger. Hence a possible objection to that work
is that the effects arising from the splittings $\Delta_{q,l}$ should not be
trusted in theories where potentially disastrous, larger effects have not
been considered. Although the considerations in the
present paper have not been specifically addressed
to the case of a unified theory, they
can nevertheless be extended to it. In this way, we think that our previous
conclusions \cite{BH} about the relevance of flavour signals in supersymmetric
unification are reinforced.


\begin{thebibliography}{99}
\frenchspacing


\bibitem{Renton} P. Renton, Rapporteur talk at the Int. Symposium on Lepton and
Photon Interaction at High Energies, Beijing (August 1995), Oxford preprint
OUNP-95-20 (1995).

\bibitem{DG} S. Dimopoulos and H. Georgi, {\it Nucl. Phys.} {\bf B193}
(1981) 150.

\bibitem{EN} J. Ellis and D. Nanopoulos, {\it Phys. Lett.} {\bf B110} (1982)
44;
R. Barbieri and R. Gatto, {\it Phys. Lett.} {\bf B110} (1982) 211.

\bibitem{GIM} S. Glashow, J. Iliopoulos and L. Maiani, {\it Phys. Rev.}
{\bf D2} (1970)1285.

\bibitem{DLK} M. Dine, R. Leigh and A. Kagan, {\it Phys. Rev.} {\bf D48}
(1993) 4269; Y. Nir and N. Seiberg,  {\it Phys. Lett.} {\bf B309} (1993) 337.

\bibitem{FlSym} For attempts at using flavour symmetries,
continuous or discrete, gauged or ungauged, Abelian or non-Abelian,
to constrain both the fermion and the sfermion spectrum, see:
Reference \cite{DLK} and
M. Leurer, Y. Nir and N. Seiberg, {\it Nucl. Phys.} {\bf B398} (1993) 319.
P. Pouliot and N. Seiberg, {\it Phys Lett.} {\bf B318} (1993) 169;
D. Kaplan and M. Schmaltz, {\it Phys. Rev.} {\bf D49} (1994) 3741;
A. Pomarol and D. Tommasini, preprint CERN-TH/95-207.
L. Hall and H. Murayama, preprint UCB-PTH-95-29.

\bibitem{BH} R. Barbieri and L. Hall, {\it Phys. Lett.} {\bf B338} (1994) 212;
R. Barbieri, L. Hall and A. Strumia, {\it Nucl. Phys.} {\bf B445} (1995) 219;
R. Barbieri, L. Hall and A. Strumia, {\it Nucl. Phys.} {\bf B449} (1995) 437.

\bibitem{PT} See A. Pomarol and D. Tommasini, reference \cite{FlSym}.

\bibitem{DH} S. Dimopoulos and L. J. Hall,
{\it Phys. Lett.} {\bf B344} (1995) 185.

\bibitem{DHR} S. Dimopoulos, L. J. Hall and S. Raby, {\it Phys. Rev. Lett.}
{\bf 68} (1992) 1984; {\it Phys. Rev.} {\bf D45} (1992) 4192;
L. Hall and A. Rasin, {\it Phys. Lett.} {\bf B315} (1993) 164.

\bibitem{GL} J. Gasser and H. Leutwyler, {\it Phys. Rep.} {\bf 87} (1982)77

\bibitem{L} H. Leutwyler, {\it Nucl. Phys.} {\bf B337} (1990) 108.

\bibitem{RPP} Review of Particle Properties, {\it Phys. Rev.} {\bf D50} (1994)
1316.

\bibitem{BFS} R. Barbieri, S. Ferrara and C. Savoy, {\it Phys. Lett.} {\bf
B110} (1982) 343;
A. Chamseddine, R. Arnowitt and P. Nath, {\it Phys. Rev. Lett.} 49 (1982) 970;
L. Hall, J. Lykken and S. Weinberg, {\it Phys. Rev.}  {\bf D27} (1983) 2359;
H. Nilles, M. Srednicki and D. Wyler, {\it Phys. Lett.} {\bf B120} (1983) 346.

\bibitem{HKR} L. J. Hall, V. A. Kostelecky and S. Raby, {\it Nucl. Phys.} {\bf
B267} (1986) 415; H. Georgi, {\it Phys. Lett.} {\bf B169} (1986) 231.

\bibitem{FN} C. D. Froggatt and H. B. Nielsen, {\it Nucl. Phys.} {\bf B147}
(1979) 277.

\bibitem{DP} S. Dimopoulos and A. Pomarol, {\it Phys. Lett.} {\bf B353} (1995)
222.

\end{thebibliography}
\end{document}